\documentclass{article}    

\usepackage{graphicx}

\title{\textbf{Satisfying Reality}}  
\author{Richard A Mould\footnote{Department of Physics and Astronomy, State University of New York, Stony Brook,
\mbox{New York} 11794-3800; http://nuclear.physics.sunysb.edu/\~{}mould}}  
\date{State University of New York, Stony Brook, N.Y.}    
   
\begin{document}           

\maketitle              

\begin{abstract}

It is claimed elsewhere that the conscious states of humans must have evolved together with their biological states,
and that an ongoing interaction between the two must have occurred to insure that they mirror one another in any
species. A quantum mechanical mechanism and an evolutionary model for the assumed mind/body interaction are
described in those papers.  The present paper outlines the related ontological and epistemological assumptions,
showing how the claimed connection between conscious states and physical states should be understood.     	

\end{abstract}    

\part{}

My basic ontological assumption is that the universe is monistic.  It is a fully unified whole that includes all
that we call objective reality, and all that we call subjective reality, where there is no fundamental distinction
between the two at this level.  Moreover, I assume that there are \emph{no} differences that are of fundamental or
intrinsic importance to the monistic universe.  There are of course differences, accounting for the great variety of
things that we experience.  Among these we find `significant' differences, accounting for the fact that some
become our guideposts for the rest.  We find \emph{conserved} quantities, (i.e., things that remain unchanged in
time), and \emph{invariant} quantities (i.e., things that remain the same under various kinds of displacement or
transformation), and these become the constants that stabilize our lives.  However, I do not think there is anything
special about these quantities in the underlying monistic universe.  They are special only to the
part of nature that they themselves circumscribe.  I therefore imagine that the fully inclusive universe is a
seamless whole that contains all discernable differences, where none are intrinsic to the universe itself.  

The above is a non-verifiable philosophical statement.  An unfettered view of the monistic universe is not possible
for humans, inasmuch as our knowledge is always based on significant distinctions of one kind or another. 
Experientially, we begin with highly selective images that comprise the foreground of our attention.  We can
certainly extend our imaginations beyond these specific gestalts by thinking logically about their content.  But
then, we become tied to a thinking process that relies on distinctions of a different kind - namely, those that
elevate and reify the primitives and axioms of our logical construction.  It is this property of ourselves in
relation to the universe as a whole that prevents us from fully grasping it as a whole.  The monistic universe has
no capital landmarks; and so, we do not have a language, and will never have a language (mathematical or otherwise)
to talk about it in its entirety.  It's as though we can see images of many kinds on a broad universal canvas, but
the unifying canvas itself is not distinguishable in a way that makes it available in one piece to our intellect. 
The existence of such a canvas is therefore an unobservable and unprovable philosophical hypothesis.  I nonetheless
believe that it describes the final nature of our own universe.		

\section*{Framented Human Knowledge}

It follows that those parts of the universe that we can intellectually grasp are always fragmented and incomplete
segments of the whole.  Let these fragments be represented by areas such as A, B, and C in fig.\ 1, where the
boundaries that separate one fragment from another are believed by us to be intrinsic separations. 
Important breakthroughs in our knowledge occur when, through some insight, the boundary between two fragments is
removed or made irrelevant.  For example, the boundary between A and B might be removed, making a larger fragment
AB.  Our knowledge would then be more inclusive.  The boundary, which was previously seen to be an intrinsic
difference between A and B, would continue to be a discernable difference, but it would no longer be intrinsic.  It
appears therefore that increasing inclusiveness and eliminating apparent intrinsic distinctions go hand in hand.  Of
course, the area AB would still be contained within a wider boundary, but the knowledge represented by this union
would be more profound. 

\begin{figure}[t]
\centering
\includegraphics[scale=0.8]{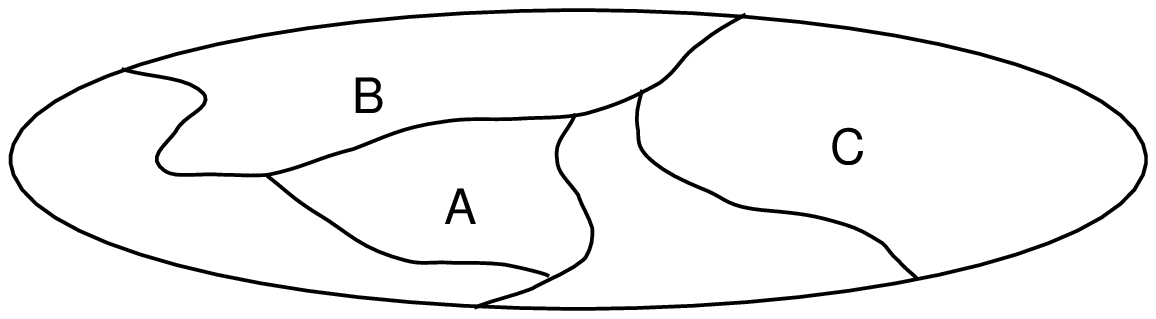}
\center{Figure 1}
\end{figure}
  
Furthermore, there will always be a wider boundary.  However many boundaries are transcended by new insights, it is
the nature of human knowledge that there will always be new absolutes to articulate any new perspective. 
Experientially, a wider and more inclusive gestalt will always have a background that is not in the picture. 
Logically, a wider and more inclusive formalism always begins with newly proclaimed axioms and primitive terms that
define the absolute or intrinsic boundaries of the new perspective.  G\"{o}del's  undecidability theorem assures us
that the new formalism, however general, cannot contain all of the true theorems in a universe of any complexity. 
Therefore, our knowledge, either experiential or formal, can never include the universe as a whole.  Beyond this
purely epistemological point, my ontological belief goes a step further to maintain that our monistic universe has
no absolute boundaries of its own, apart from our inability to discover them. 

\section*{Splitting the Whole}

However satisfying or unsatisfying a monistic universe might be as a philosophical premise, it cannot serve as the
basis of a usable ontology. We cannot launch an articulation of reality from a platform that is ineffable in
principle.  To proceed, we must create some arbitrary boundaries.  I therefore divide the monistic universe into
three categories: \emph{matter}, \emph{form}, and \emph{consciousness} (see fig.\ 2).  These, I believe, are the most
significant ontological divisions that can usefully serve as a basis of human understanding.  Penrose split the
universe up in a similar way\cite{RP}.  In addition, the grouping of these ontological categories in fig.\ 2 allows
a further epistemological division between the subjective world and the objective world.  This is the Cartesian
divide, and is essential to what follows.

	The objective world in fig.\ 2 is a combination of matter and form.  This is a part of the universe that is not
directly accessible to our selves.  The subjective world in fig.\ 2 is a combination of form and consciousness.  This
is the part of the universe that is directly accessible to our selves.  Form is the common element.  It is the link
that we conscious beings have to the material world around us.  We have no direct knowledge of matter or the form
that it takes; however, we do have a direct knowledge of consciousness and the form that it takes. Our basic
epistemological assumption is that a correspondence can be established between the form that matter takes, and the
form that consciousness takes.  

Matter, form, and consciousness are such primitive ontological ideas that it is not possible to give them
definitions independent of context.  Like the primitives of a logical system, their meaning can only be derived from
the way that they are used.  However, an indication of intent is possible.  I will say that \emph{consciousness}
includes perceptions such as sight, sound, taste, etc., and emotions such as fear, anger, and love, etc.  I call
these the elements of `pure consciousness' when they are experienced in isolation.  More generally, consciousness
takes on a variety of \emph{forms} when complex images and ideas are held in mind.  It is through these images and
ideas that we strive to portray the properties of the objective world. 

The early pre-conscious universe of cosmology consisted of \emph{matter} in one form or another.  It does not make
sense to imagine that the universe at that time possessed form alone. There had to be something that assumed the
many forms that we study in cosmology, and I call that something matter.  The form that matter takes is the subject
of all of the physical sciences.  I will also refer in this paper to pure matter, or \emph{formless matter}, whose
properties are not expressible in any formally structured science.  I cannot explain what I mean by this until some
further issues are clarified, and the idea is put to use in Part II.  

	The divisions in fig. 2 are arbitrary in the sense that they are unsupported by the monistic universe itself. 
Nonetheless, they are virtually inevitable to any notion of reality that acknowledges the existence of a world that
lies beyond our separate selves, and is common to our many selves.  This necessitates the recognition of two realms
of reality, and a connecting link between them through which we can transcend ourselves to find the universe we
share.  

\begin{figure}[t]
\centering
\includegraphics[scale=0.8]{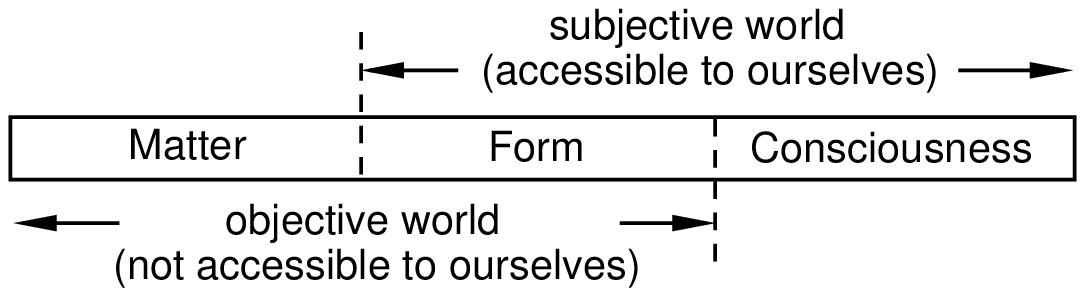}
\center{Figure 2}
\end{figure}

\section*{Subjective Formations}

	Beyond the emotions and bare sensations that pervade our lives, our experience is populated with many highly
developed \emph{images} of the things that we find in our subjective environment, such as tables, chairs, wine
bottles, etc.  These images have their origin in ordinary perception.  By logical extension, we also constructed
images of things that we do not (directly) experience, such as the atoms and molecules that are assumed to make up
those tables and chairs and bottles of wine.  Since these are not directly seen, they do not have their origin in
ordinary perception.  Instead, \emph{constructed images} like these are formed in our imagination, in this case, in
our scientific imagination. A third kind of subjective experience is an \emph{idea}.  This is formed when we discover
a relationship between images and/or parts of images, or between constructed images and/or their parts.  

These three formations are not clearly distinct from one another.  Most images, and certainly those of the
constructed variety, are permeated with ideas.   However, it will serve our purpose in the next section to
distinguish between things such as wine bottles (images), and their molecules and atoms (constructed images), and
the physical laws (ideas) that govern their behavior.

\section*{Epistemology out of Ontology}
  
The relationship between these subjective formations and the ontological categories in fig.\ 2 is shown
diagrammatically in fig.\ 3.  \emph{Subjective images and ideas} are represented there by the three rectangles on the
top row.  Each is assumed to correspond to something equivalent to itself in the objective world, where these
\emph{objective things and relationships} (bottom row) are joined to their subjective counterparts by double lines
called \emph{rules of correspondence}.  We are the ones who make the rules of correspondence, by virtue of the
relationship that we assume exists between the objective world and our introspections about it\cite{HM}.  They
connect the form that consciousness takes with the presumed form that matter takes.  

Thus, I believe that our subjective concept of a bottle of wine (i.e., the cross-hatched rectangle at the top-left
in fig.\ 3) corresponds to some such thing in the objective world (i.e., the cross-hatched rectangle at the
bottom-left).  I also believe that the subjective concept of an atom (i.e., the broken-cross-hatched rectangle at
the top-center) corresponds to something having those same formal properties in the real world (i.e., the
broken-cross-hatched rectangle at the bottom-center). And I believe that our idea of momentum conservation (i.e.,
the gray rectangle at the top-right) corresponds to a law of nature (the gray rectangle at the bottom-right). The
space on the top-right in fig. 3 is intended to represent pure (formless) consciousness, and the space on the
bottom-left is pure (formless) matter.   

\begin{figure}[t]
\centering
\includegraphics[scale=0.8]{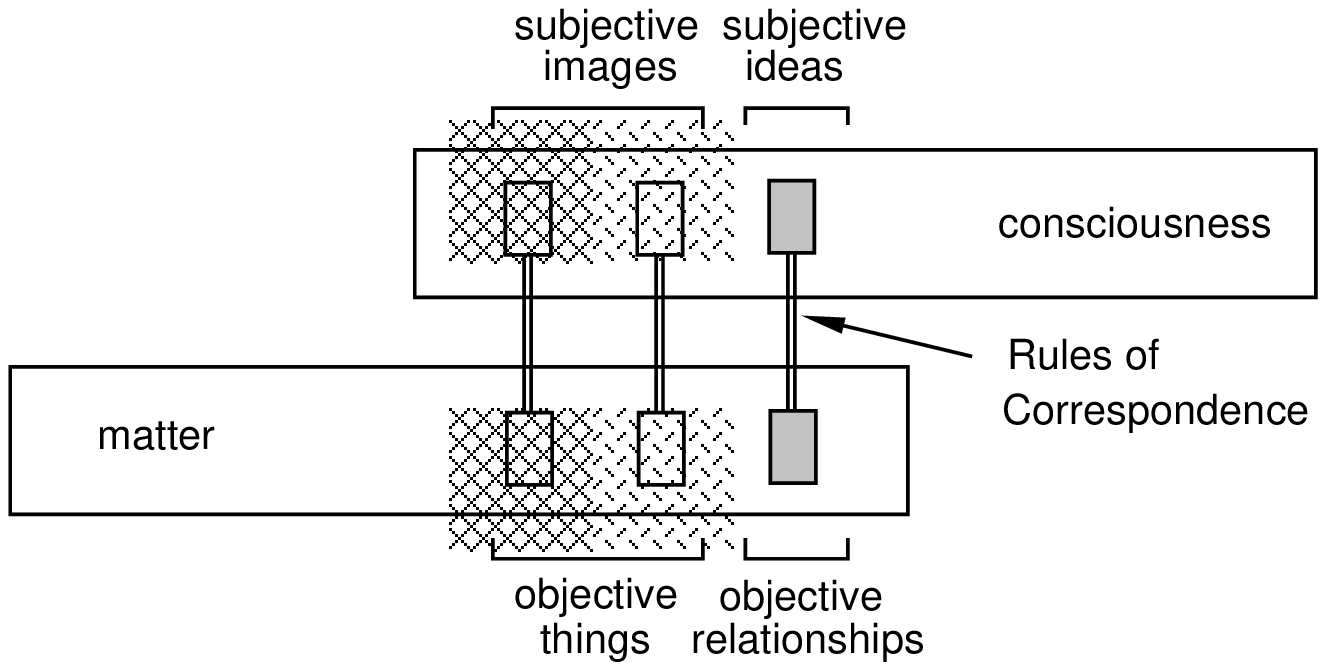}
\center{Figure 3}
\end{figure}

My subjective image of a wine bottle might be an experience in black and white, or it might include a colorful
label.  However, it is only the formal features of any image (e.g., its shape) that can be set into correspondence
with the objective thing.  All of the characteristics of pure consciousness that Galileo would have called
``secondary qualities" (e.g., its color), have no meaning in the objective world.  

\section*{Adequacy}

The subjective world in fig.\ 2 is not intended to be that of any particular individual.  Individuals are not
important here.  If no conscious individuals existed, then of course subjectivity would not exist.  But we split the
universe into three parts, not 6 billion or more.  Presumably there are as many images of a wine bottle in the
universe as there are conscious wine drinking individuals.  Each of these images will be given its own cross-hatched
rectangle in fig.\ 3, and hopefully, each will correspond to an actual wine bottle in the objective universe.  I
believe that each of the image/thing or idea/relationship pairs in fig.\ 3 represents an attempt on the part of the
universe to understand itself.  It is ironic that a monistic universe should have trouble with such a seemingly
dualistic task - that of understanding itself.  However, a failure to match pairs in fig.\ 3 does not constitute a
contradiction on the part of the monistic universe, for there is no reason why every subjective image that the
universe creates should correspond to something in the objective part of itself.  No philosophical wine drinker can
be absolutely sure that the table, the chair, or the bottle he holds is as objectively real as it appears.  It is
for this reason that we must continuously test our subjective images and ideas for their adequacy in portraying an
objective world.  Doing so is necessary to the long-term survival of a conscious species, and the survival of such a
species is necessary if the universe is ever to understand itself.   

\part{}

The Ontological/Epistemological scheme in Part I will now be extended in two ways.  The first gives a more accurate
account of the way that physicists introduce matter into theoretical systems, and the second anticipates a possible
two-way involvement of consciousness with objective physical systems.

\section*{Rules of Stipulation}

Newton's physics may tell us how billiard balls behave, but it does not tell us when and where to find billiard
balls.  Therefore, when defining a physical system, a physicist must \emph{stipulate} when and where material objects
exist.  He quantifies the initial state of a billiard ball by giving its position, size, mass, linear and angular
momentum.  The same kind of thing must be done when specifying the initial state of an elementary particle in a
quantum mechanical system, but in this case, the \emph{rules of stipulation} are different.  The rules in this case
require one to fix the variables of a quantum mechanical state function.  For an electron that means giving its mass,
charge, spin, and its probability amplitude in configuration and momentum space.

The checkered rectangle on the top of fig.\ 4 represents any subjective image or idea that includes a piece of
matter. The objective world in that figure contains the corresponding piece of matter in question, but we have
unnaturally divided that world into matter and form.  Pure matter, represented by the black square on the lower
left, is therefore unnaturally separated from the form that it takes, which is represented by the checkered
rectangle in the lower center of the figure. We repair this separation by connecting the two things with
double lines representing \emph{natural associations}. The reason for this diagrammatic separation is given in the
final paragraphs of this section. 

The rules of stipulation are shown entering from the upper left in fig.\ 4.  They are intended to be parallel to the
natural associations of the objective world; but of course, they cannot terminate on anything to the left because
they do not complete an association of any kind.  They only appear in current theories to announce that a given state
of matter exists at a certain time and place in our theoretical system.  The declaration itself must be stated in
the formal language of the theory, even though it represents missing information about pure matter.  Figure 4
therefore embodies an amendment to our epistemological/ontological scheme that describes how matter is formally
introduced into theoretical physics.  

You might imagine that physics will one day have a theory that automatically includes a complete state description
of all particles at any time in the history of the universe.  That would seem to make rules of stipulation
unnecessary.  However, quantum mechanics can only predict probabilities.  This means that an extra-theoretical rule
of stipulation will always be necessary \emph{after any measurement} in order to specify which of the possible
results of the measurement is realized.  Therefore, rules of stipulation are necessary to say how quantum mechanical
particles survive measurement.  

\begin{figure}[t]
\centering
\includegraphics[scale=0.8]{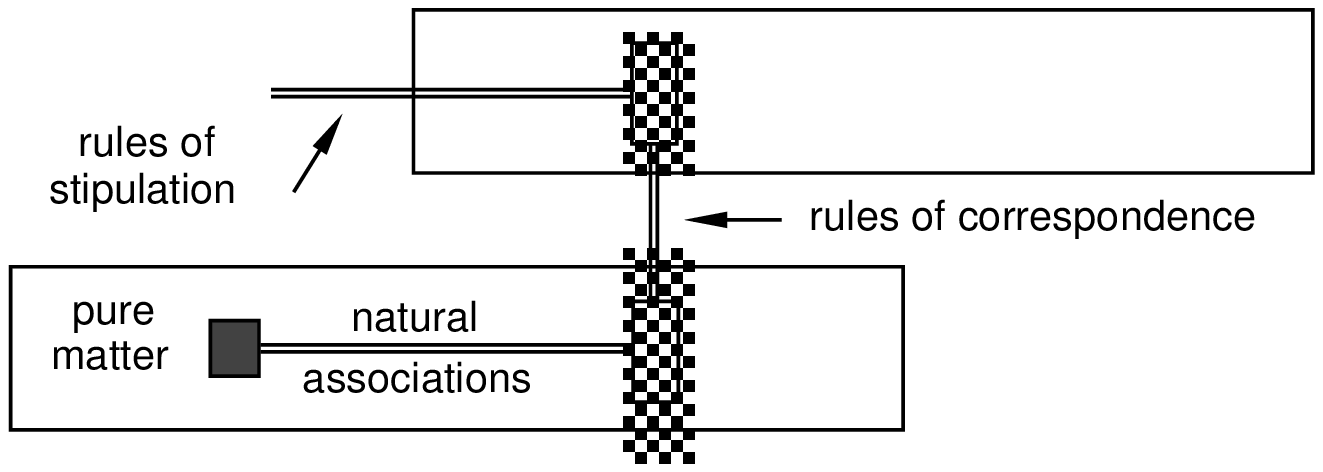}
\center{Figure 4}
\end{figure}

Here finally is the reason I give for the possibility of pure formless matter, and the diagrammatic separation within
the objective world in fig.\ 4.  Formalists and/or subjectivists might have objected to our ontological scheme from
the beginning, claiming that the `matter' category in fig.\ 2 serves no useful purpose; and that therefore, we who
are confined to conscious awareness can get along without it.  Surely they will say that the `pure matter' referred
to in fig. 4 is a fiction.  However, one lesson of quantum mechanics is that there is no deterministic \emph{form}
that allows us to predict everything about matter that can be measured.  We can predict probabilities over ensembles
of measurements, but we must resort to extra-theoretical rules of stipulation in order to specify the results of
individual measurements.  This need for rules of stipulation is a sign that `form' fails to capture and embody every
part of nature.  Form succeeds in capturing and embodying ensembles of measurements, but strangely enough it fails
at the level of the individual.  Therefore, so long as we need rules of stipulation, the natural associations that
parallel the rules of stipulation will serve the purpose of connecting that part of the objective world that can
manifest itself in a predictable form, with the part that cannot.    

Of course, we might one day abandon quantum mechanics and go back to a determinism that allows the results of a
measurement to be known, or at least knowable, in advance.  Such a theory might conceivably be so powerful that all
rules of stipulation will be formally included in it.  If that happens, then one might reasonably conclude that a
separate ontological category of pure matter is not necessary.  However, I do not believe that this will happen, for
it is unlikely that classical determinism can be convincingly revived.  Therefore, as matters now stand, matter
stands apart from consciousness and form.

\section*{The Consciousness Connection}

It is generally assumed that natural physiological processes within the body give rise to consciousness, even
though no one can say how or why something like that should occur.  Scientists generally believe that this
consciousness is epiphenomenal.  That is, they believe that the body can create and influence consciousness but
that consciousness cannot influence the body.  Accordingly, the mind/body influence is believed to be a one-way
street.  

William James challenged the epiphenomenal idea, saying that consciousness and matter must `interact' with one
another in order to satisfy the requirements of subjective evolution.\cite{WJ}   He accepts the psycho-physical
parallelism of von Neumann\cite{JvN}, which I interpret as the parallel relationship that exists between the various
subjective and objective things in figs.\ 3 and 4.  James says that such a parallelism would not be possible if the
subjective and the biological states of a species did not actively engage one another during the time of their
evolution.  In particular, subjective states must have a consequence for biological states if there is to be
Darwinian selection against a ``wrong" subjective construction.  Without feedback of this kind, subjective imagery
would have developed independently of the underlying biology, so there would have been no evolutionary mechanism to
keep subjectivity on a parallel course with objective reality.  \emph{In these circumstances, the final emergence of
a true parallelism would be a miracle}. Therefore, if we are to reject miracles, and accept the idea that
subjectivity evolved along with everything else, and if we are to avoid a vacuous Berkeleian idealism, then we must
accept von Neumann's psycho-physical parallelism and acknowledge its origins in a Jamesean-like evolutionary
feedback.  It is for physics to discover the nature of this feedback and incorporate it into physical
theory\footnote{The psycho-physical parallelism of von Neumann resembles the pre-established harmony of Leibniz. 
For Leibniz, this harmony of correspondences between subjectivity and objectivity is indeed a miracle that is
arranged by God.  However, I say that it is a result of natural `monistic' processes that find their way into human
consciousness through the evolutionary mechanisms of natural selection.}. 	

I accept and extend the argument from evolution in some recent papers, and speculate as to how
a ``consciousness-to-matter" influence might have come about. \cite{RM98,RM99}.  But apart from the particular
mechanism of this influence, or the success of any particular speculation about it, we must provide for its
epistemological possibility.   The scheme in fig.\ 3 does not allow such an influence to take place in either
direction.  It portrays the objective world as a self-contained automaton that makes no authentic connection to the
conscious world, except for rules of correspondence that may or may not work.  A further modification of fig.\ 3 is
therefore required.  

Imagine that the brick-faced rectangle in the objective world of fig.\ 5 represents the body of a person who we call
``Harry", and who we assume has a conscious life.  In a monistic universe, Harry's consciousness would not be
artificially separated from his body; so his consciousness is related to his objective body in fig.\ 5 through
\emph{natural stipulations}, represented by the double line that comes into the figure from the lower right.  This
does not terminate on anything on its right because Harry's associated consciousness is not included in the
diagram.  It is my belief that the stipulated influence goes both ways.  Harry's objective body exerts an influence
on his consciousness (again, not in the diagram), and his consciousness exerts an influence on his objective body,
as per the argument of William James. The mind/body interaction occurs here.

\begin{figure}[t]
\centering
\includegraphics[scale=0.8]{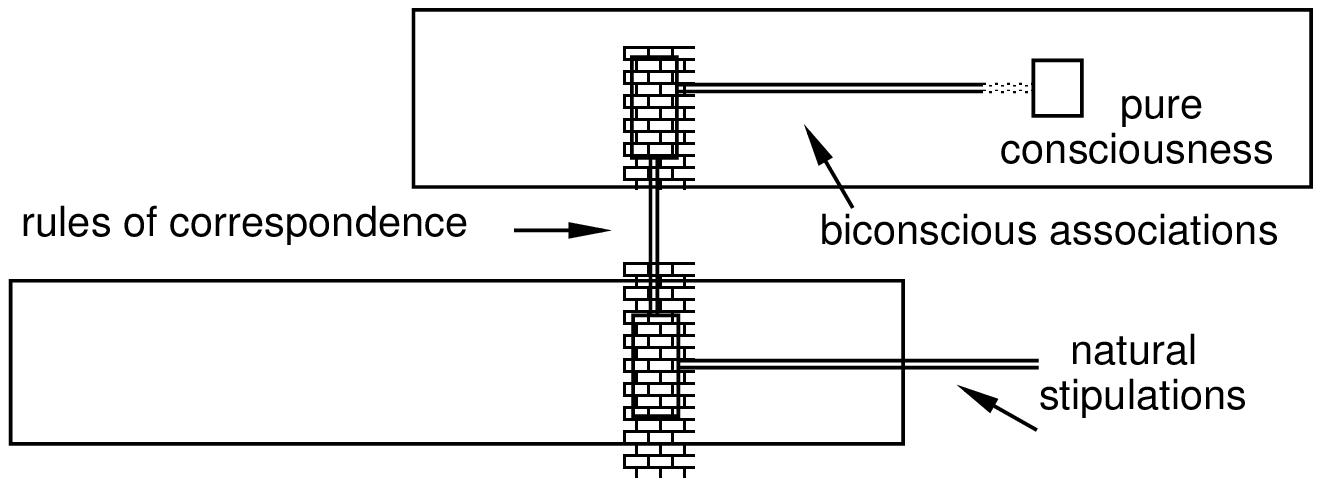}
\center{Figure 5}
\end{figure}

	 The brick-faced rectangle on top in fig.\ 5 represents an image of Harry's body in the mind of an \emph{external
investigator} who is giving serious thought to Harry's body and Harry's experiences.  The consciousness that the
investigator associates with Harry's body is represented by the small white square on the upper right, which is also
in the mind of the investigator.  Therefore, the upper brick-faced rectangle plus the white square
constitute the investigator's model of Harry's mind \& body.  The model cannot show an intrinsic connection between
these two things, so they are joined instead by \emph{biconscious associations} that connect specific parts of
Harry's (theoretical) physiology to specific kinds of consciousness.   These associations are ``biconscious" because
they paste together two alien things that are in the mind of the investigator: (1) parts of his physiological model,
with (2) elements from his pure conscious awareness.  

For instance, if the investigator believes that the color `red' is associated with neurological events at a point in
Harry's prefrontal cortex, then he will make a biconscious association that goes from the prefrontal cortex of his
physical model (top-center in fig.\ 5) to his own concept of pure red consciousness (top-right).  An arranged
wedding of red-with-brain is necessary because redness cannot be included in a formal theory of the kind entertained
by theoretical physics, or theoretical physiology.  This reflects a separation between form and consciousness that
occurs within the subjective world.  Biconscious associations are intended to run parallel to the natural
stipulations entering the objective world in fig.\ 5, fulfilling the requirements of von Neumann's psycho-physical
parallelism.  In the least, they parallel the one-way (epiphenomenal) influence of physiology on psychic states;
but they also parallel the proposed two-way influence.  

\section*{Super Theory}

If a biconscious association is verified, it will be an empirical relationship that cannot be integrated into
theoretical physics as that science is now practiced and understood.  That's because physical (or physiological)
variables and psychological states are not commensurable.  Physiological variables are connected to the objective
world \emph{via} vertical rules of correspondence; whereas psychological states exist in another part of the
subjective world that is related to theoretical physiology \emph{via} horizontal biconscious associations. Of course,
psychological states might some day be expressed as psychological variables that are related in a definite way to
physiological variables; however, relationships like that do not presently exist.  At the present state of our
scientific understanding, consciousness can \emph{only} be joined to physiology through biconscious associations; and
this means that the connection cannot be organic.

It is possible in principle to find a \emph{theory of consciousness} concerning relation-ships between psychological
states.  Presumably, such a theory would employ psychological variables that represent states of consciousness such
as `redness'.  These variables would then be related to their intended psychic states \emph{via} rules of
correspondence, not by biconscious associations\footnote{Correspondences are much stronger than associations.  A
psychic state can only be loosely associated with a physiological configuration, whereas it can be set into full
correspondence with a theoretical variable that is intended to represent it in some (now non-existent) theory.}.  A
theory of consciousness would therefore employ horizontal rules of correspondence that go from a theoretical
construction like the upper brick-faced rectangle in fig.\ 5, to the conscious state that it represents like the
white square in fig.\ 5.   

	Now imagine a \emph{super theory} that combines our hypothetical theory of consciousness with physics.  That would
be a non-trivial unification of the two theoretical systems that establishes a functional relationship
between their variables.  Super theory would therefore require both vertical and horizontal rules of
correspondence.  It should be possible to reduce such a theory to: (1) a physics part that uses physical variables
only, (2) a psychological part that uses psychological variables only, and (3) a part that establishes the functional
relationship between the two kinds of variables.  The latter would give us formal representations of the biconscious
associations.  

This would allow us to derive biconscious associations from a wider theoretical framework, and that would give us
the only logical `explanation' of their existence that we could ever expect.  When Planck discovered his radiation
law between intensity and wave length, no one concluded that wavelength thereby `caused' intensity, or that
intensity `caused' wave length, or that anything at all was `explained' by this empirical formula.  It was not until
Einstein's photon theory provided a logical context for Planck's formula that it was possible to believe that it was
thereby explained.  Only then could a causal pathway be derived from the basic theory to the Planck formula. 
Similarly, the discovery of biconscious associations should not lead one to conclude that certain physiological
configurations `cause' psychological states, or that psychological states `cause' physiological configurations, or
that anything at all is thereby `explained'.  Explanations and causal pathways will exist only if a wider theory
such as the above super theory allows one to derive the biconscious associations in question.  This will be true
independent of the possible one-way or two-way influence of these associations.

\section*{QMod}

Suppose that a modification of quantum mechanics is found that produces the collapse of a state function, \emph{and}
embodies physical mechanism that fully explains the evolution of von Neumann's psycho-physical parallelism. I will
call such a modified theory QMod.  If furthermore, the physiological site of this modification is also the site of a
subjective experience that does not have an interactive influence on QMod, then it would seem that an epiphenomenal
theory of consciousness is again indicated.  In other words, if QMod can account for a state reduction and the
psycho-physical parallelism in a self consistent and self contained way without making explicit use of conscious
states, then consciousness would again appear to be a non-participant as it has been since the time of Newton. 
However, QMod suffers the same disadvantage as any epiphenomenal theory.  It may be able to explain the workings of
a material system without the help of consciousness, but it cannot explain the co-appearance of consciousness
itself.  Only a super theory can hope to explain the conscious states that are associated with a material system.  If
that happens, if a super theory were to prove successful, then consciousness and matter would become partners in our
understanding of nature, and an epiphenomenal interpretation would lose all significance. In the end,
epiphenomenalism remains an impoverished way of looking at QMod or any purely mechanical system that is
(tentatively) at the forefront of theoretical construction.     

		A super theory of the kind described above is far beyond anything that is currently possible, and may never be
realized at all.  Therefore, we should not try to explain consciousness at this point.  Instead, we should
concentrate on more limited objectives such as discovering the material circumstances in which consciousness
appears, and finding the biconscious associations that chronicle the evolution of the psycho-physical parallelism. 
The latter is what the author attempts to do in refs.\ 5 and 6.

\section*{Incompleteness and a TOE}

When physicists speak about a Theory of Everything (TOE), they are not concerned about including pure consciousness
or formless matter into some grand theoretical scheme.  They are only concerned with those things that are contained
within the `form' part of the universe (i.e., the overlapping subjective and objective worlds in   fig.\ 3 that are
joined with rules of correspondence.)  Presumably, such a TOE would cover the physical and biological sciences,
economics, sociology, and behavioral psychology, as well as all other `departments' of objective knowledge.  Many
physicists believe that physics already includes all of this knowledge in principle.  I take this to mean that they
believe that the laws of physics can be put into an axiomatic form such that all objective laws or relationships
that are found in nature and in human affairs are included as theorems.  I do not believe this to be possible.  One
obvious objection comes from G\"{o}del's undecidability theorem, which tells us that no finite set of axioms can
successfully derive all true theorems in a complex universe containing discrete variables.  

Certainly a TOE of the kind envisioned by physicists can be closely approached, even if it is never fully realized. 
Further unification in physics is surely possible, and is the source of great motivation in such areas as string
theory.  But human knowledge is essentially fragmented and incomplete as has been said.  However inclusive our
understanding of nature, it will always be contained within a boundary that excludes something else.  In particular,
pure consciousness and formless matter are excluded in the kind of TOE that is pursued in physics. 

The super theory postulated above represents an even wider theoretical aspiration - one that breaks through the
bounds of the Cartesian divide between the objective and the subjective worlds.  It imagines a formalism that
significantly includes variables from both worlds, a feat that may never be realized.  But even if something like
this is achieved, it will surely be bounded in such a way as to exclude some other part of reality.  Again,
G\"{o}del's undecidability theorem places limitations on a formalism of any complexity that contains discrete
variables; and that stricture applies to our super theory as well as the any unified theory of physics.  The theorem
represents a purely formal limitation on any TOE or super TOE.  

But more than that, we know that super theory excludes formless matter.  As previously explained, matter must be
introduced into physics by stipulation because of the quantum mechanical unpredictability of many of its numerical
values upon measurement.  This means that no formal theory, however broad, can include that part of reality.  The
measured ``eigenvalues" of matter will always have to be stipulated \emph{extra-theoretically} so long as quantum
uncertainty is integral to physics.

\section*{The Choices}

In the best of all worlds, we humans would be able to fully comprehend the monistic universe.  However, I do not
believe that this is possible for mortals of this world.  We can certainly talk about a monistic universe and
speculate about some of its holistic properties as we have done in this paper.  But it is not possible for us to put
any particular thing (e.g., a wine bottle) into a full monistic context.  That's because the universe is not
circumscribed by intrinsic boundaries that allow us to comprehend it.  To get a handle on wine bottles and their
like, we are forced to introduce arbitrary boundaries that are useful for that purpose.  

The boundaries we introduced appear in figs.\ 2 and 3.  They first divide the universe into three ontological
categories, and then split it into two epistemological parts.  One of these parts includes that which we conscious
beings know to exist, and the other is that which we can only imagine exists.  This division of reality has a
demonstrated utility, for it has served as a basis for accumulated knowledge since the early seventeenth century. 
At the time, Descartes' separation of mind and body was put to use in Galileo's distinction between primary and
secondary qualities.  Without this distinction, science as well as most other fields of knowledge would not have
advanced much beyond their medieval state. 

There are certainly other choices that could be made.  We might try to limit ourselves to a purely subjective
ontology, as positivists are prone to do; or, we might claim that there is no significant separation between the
subjective world and the objective world, as some realists are prone to say.  Both of these choices attempt to mend
the Cartesian divide as though the monistic universe is fully available to our intellects.  However, I do not think
it is available in this way.  

Perhaps the super theory, if it were realized, would suggest another way of splitting up the universe.  Because its
rules of correspondence seem to ignore the Cartesian divide by radiating in both horizontal and vertical directions,
super theory might suggest novel ontological and/or epistemological ways of introducing distinctions into a monistic
universe.  That is certainly a possibility, although it is not one that can be seriously entertained in the absence
of a well-defined super theory.  My belief is that the boundaries in fig.\ 2 will survive in any case.  The
separation between that part of the universe that we directly know (i.e., the subjective) and the part that we can
only imagine (i.e., the objective) is much too fundamental an epistemological distinction to be cast aside. 
Descartes is likely to prevail at the epistemological level no matter what happens.  Furthermore, the ontological
distinction between pure consciousness and the form that it takes is hard to deny, not to mention the need to
recognize the existence of formless matter that is likely to remain outside of any theoretical system.  It is for
these reasons that I believe that the categories and divisions in fig.\ 2 are, and will continue to be, a practical
starting point for any scientific ontology and epistemology.   

Our theories can only be related to something else by means of either rules of correspondence, or rules of
stipulation, or biconscious associations.  I have not made a point of this, but even the subjective relationships
between our images and our constructed images are established by rules of correspondence.  For instance, only a
rule of correspondence can connect a theoretically constructed meter stick with the visual image of a meter stick
that we experience in the laboratory\footnote{Constructed images of meter sticks not only include atomic or molecular
models, but they also include the elongated rectangles that we drawn on a blackboard to help solve problems in
kinematics.  They include any cognitive attempt to represent something by abstracting from it, or by elaborating
upon it.}.  This is the original meaning of a rule of correspondence introduced by Margenau (ref.\ 2), for whom
direct experience is the ultimate reference for any theoretical activity.  In such a purely subjective context,
von Neumann's psycho-physical parallelism can only point to the similarity that exists between our images and our
theoretical constructions, and this creates a problem.  James's evolutionary logic could not have begun by mediating
between images and constructions alone, for there is no reason why either of these should have existed prior to a
time of their Darwinian emergence.  They would have served no purpose prior to that time.  Therefore, subjective
evolution could not have begun within a purely subjective ontology.  It could not have gotten started without an
external taskmaster that makes `realistic' survival demands of some sort on the evolving subjective states.

\section*{Copenhagen}

	The Copenhagen interpretation of quantum mechanics embraces a subjective philosophy, for it is claimed that we can
no longer assume the existence of an objective world.  Heisenberg's uncertainty principle is often cited as the
reason for this confinement to subjectivity.  That principle tells us that there is something incomprehensible about
nature because we cannot express the theoretical state of a physical system in terms of classical variables to any
degree of precision.  We are forced instead to introduce \emph{probability} as an intrinsic property of the system. 
It is this rather strange feature of quantum mechanics that is said to refute the idea of `objectivity'.  However,
it only refutes the idea of `classical objectivity'.

	From the beginning of this paper I have accepted that there is something incomprehensible about nature.  I believe
that we humans can only go so far in understanding the whole.  But that has not prevented me from positing the
existence of an objective world, which ``lies beyond our separate selves, and is common to our many selves".  Nor
does it prevent me from creating theoretical models that are consistently related \emph{via} rules of correspondence
to something that I believe to have similar formal properties in the objective world.  To this end, one need only
accept \emph{intrinsic probability} as a primitive variable in quantum mechanical systems.  It is clear, therefore,
that Copenhagen physicists were not forced to give up the objective world.  They \emph{chose} to do so, citing the
demise of classical objectivity as their reason.  

	If quantum mechanics tells us anything that is epistemologically novel, it is that we will always need ``rules of
stipulation" to specify the state of a particle that survives quantum measurement.  These are extra-theoretical
statements that specify the actual measured values of intrinsic probability distributions.  A Copenhagen theorist
would say that these stipulations are the principal evidence of the claimed incomprehensibility of nature.  I agree,
but I would put it differently.  I cite these rules as evidence that there is a part of the objective world (namely,
pure-formless matter) that cannot be included in our theoretical system. Only the `form' that matter takes can be
theoretically included.  Rules of stipulation therefore point to a part of nature that we cannot describe.  They do
not say that there is no other part of nature.  

And finally, as stated in the previous section, a completely subjective ontology cannot give an evolutionary account
of its own emergence.  There is no place for bare subjectivity to begin.  It would have to be born whole. 
Consciousness cannot emerge in bits and pieces, without another (external) reality to mediate the process -
selecting pieces that fit into a survival pattern of the species, and discarding pieces that don't fit.  Evolution
implies building piece by piece in this way, and biological evolution requires an unforgiving environment to be the
womb of this process.  Psychological evolution requires no less.  With the Copenhagen approach, as with all
subjectivist philosophies, subjective states lack an adversarial environment of this kind to rub against.  And this
implies the miraculous emergence of a full-blown and fully consistent psychic life, or possibly, a more gradual
unfolding of serendipitous match-making between the bits and pieces of psychic life.      

\section*{Revealing Universals}

For one reason or another, I believe that we will never have a fully complete theory of everything.  It is probably
an insight like this that has led Post Modernists and others to abandon the quest for a common universal knowledge
and a common universal understanding, choosing instead to go down a separatist path.  If human knowledge is
fragmented and incomplete, these philosophers invite you to choose an incomplete fragment that best fits your
temperament, and become a partisan of your own truth and your own reality.  Not only is this advice drearily
contentious, but it doesn't follow.  It is true that any reality we choose will be an incomplete fragment of the
whole.  But not all fragments are equal.  There is a highly principled choice between broad-based fragments that
reveal universals, and those fragments that are narrowly provincial and/or merely self-serving.  Only the broadest
principles of unification can bring us close to the common canvas that underlies all of our diverse human
experiences.  We may never apprehend that canvas in an absolute sense, but the closer we approach it, the greater is
our reward.  Those insights that are most inclusive and least dependent on intrinsic difference are those that will
bring us the greatest intellectual and esthetic appreciation of things.  Not only do they lead away from partisan
disputation, but they stress the universals that most nearly and most beautifully reflect the underlying harmony of
our universe.

\end{document}